\documentclass[twocolumn,showpacs,preprintnumbers,amsmath,amssymb,floatfix]{revtex4}
\usepackage[utf8]{inputenc}
\usepackage[caption=false]{subfig}
\usepackage{graphicx}
\usepackage{amsthm}
\usepackage[overload]{empheq}

\usepackage[colorlinks=true, linkcolor=blue, bookmarks=true]{hyperref}

\newcommand{\comment}[1]{}

\theoremstyle{remark}

\begin{document}
\title{Schr\"{o}dinger--Newton--Hooke system in higher dimensions.\\ Part I: Stationary states}
\author{Filip Ficek}
 \email{filip.ficek@doctoral.uj.edu.pl}
\affiliation{
Institute of Theoretical Physics, Jagiellonian University, \L{}ojasiewicza 11, 30-348 Krak\'{o}w, Poland
}

\date{\today}

\begin{abstract}
The Schr\"odinger equation with a harmonic potential coupled to the Poisson equation, called the Schr\"{o}dinger--Newton--Hooke (SNH) system, has been considered in a variety of physical contexts, ranging from quantum mechanics to general relativity.  Our work is directly motivated by the fact  that the SNH system describes the nonrelativistic limit of the Einstein-massive-scalar system with negative cosmological constant. With this paper we begin the investigations aiming at understanding solutions of the SNH system in the energy supercritical spatial dimensions $d\geq 7$, where we expect to observe interesting short wavelength behaviours due to the confinement of waves by the trapping potential. Here we study stationary solutions and prove existence of one-parameter families of nonlinear ground and excited states. The frequency of the ground state as the function of the central density is shown to exhibit different qualitative behaviours in dimensions $7\leq d\leq 15$ and $d\geq 16$, which is expected to affect the stability properties of the ground states in these dimensions. Our results bear many similarities to the analogous problem that has been studied for the Gross-Pitaevskii equation.
\end{abstract}


\maketitle

\section{Introduction}
In this paper we consider the system
\begin{subequations}\label{eqn:SNH}
\begin{align}[left ={ \empheqlbrace}]
 i\partial_t \psi &=-\Delta \psi+\Omega^2 |x|^2 \psi+\psi v,\label{eqn:SNHa}\\
\Delta v &= |\psi|^2, \label{eqn:SNHb}
\end{align}
\end{subequations}
for a complex-valued function $\psi(t,x)$ and a real-valued function $v(t,x)$, where $x\in\mathbb{R}^d$, and $\Omega$ is a number. Solving the Poisson equation \eqref{eqn:SNHb} using the Green function for the Laplace operator and substituting the result into Eq.\ \eqref{eqn:SNHa} yields the Hartree-type  equation 
\begin{align}\label{eqn:SNHt}
i\partial_t \psi &= -\Delta\psi+\Omega^2|x|^2\psi-A_d \left(\int_{\mathbb{R}^d} \frac{|\psi(t,y)|^2}{|x-y|^{d-2}}\, dy\right) \psi,
\end{align}
where $A_d=\Gamma(d/2)/2\pi^{d/2}(d-2)$.  Following the tongue-in-cheek terminology of \cite{Biz18}, we shall refer to Eq.\ \eqref{eqn:SNHt} as the Schr\"odinger-Newton-Hooke (SNH) equation. This equation has been considered in three dimensions as a mean-field limit of a non-relativistic bosonic system with two-body interactions, confined in a harmonic trap \cite{Fro03}; see also \cite{Car05, Che17, Che18}.

Our interest in the SNH equation is motivated by the fact that it arises  as a nonrelativistic limit of the Einstein-Klein-Gordon system with negative cosmological constant $\Lambda$ \cite{Biz18}. The consistency of this limit requires the product $-\Lambda c^2$ to approach a positive constant $\Omega^2$ as the speed of light $c\rightarrow \infty$, yielding the coefficient of the harmonic potential in Eq.\ \eqref{eqn:SNHt}. Thus, the confinement of waves in asymptotically anti-de Sitter (AdS) spacetimes due to the gravitational potential
 translates in the nonrelativistic limit to the trapping by the harmonic potential.  From this perspective,  it is interesting to see whether solutions of the SNH equation  exhibit a behavior analogous to the instability of the AdS spacetime \cite{Biz14}. We remark in passing that the corresponding nonrelativistic limit of the Einstein-Klein-Gordon system with zero cosmological constant (i.e., for asymptotically flat rather than asymptotically AdS spacetimes), resulting in the Schr\"odinger-Newton (SN) equation (i.e., Eq.\ \eqref{eqn:SNHt} with $\Omega=0$), has been considered (under the names of the Hartree, Schr\"odinger-Poisson, or Choquard equation) in various physical contexts, for example  in modelling boson stars \cite{Kau68, Ruf69}, in attempts to envisage the wave function collapse as a gravitational phenomenon \cite{Pen96,Mor98}, and as a mean-field approximation for many-body problems; see \cite{Mor17} for a review.
 
For physical reasons, most of the above mentioned studies were restricted to three spatial  dimensions, however from the mathematical and AdS related viewpoints it is interesting to consider the SNH equation  in higher dimensions, in particular for $d\geq 6$. 
To see why $d=6$ is distinguished,  let us recall that the SNH equation preserves the mass and energy defined respectively as
\begin{align*}
M(\psi)=& \int |\psi|^2\,dx,\\
E(\psi) =& \frac{1}{2} \int |\nabla \psi|^2\,dx + \frac{\Omega^2}{2} \int |\mathbf{x}|^2 |\psi|^2\,dx\nonumber\\
&-\frac{A_d}{4} \int \left(\int \frac{|\psi(t,y)|^2}{|x-y|^{d-2}}\, dy\right) |\psi|^2\,dx\,.
\end{align*}
The SN equation enjoys the scaling symmetry
\begin{align*}
\psi(t,x)\mapsto \psi_{\lambda}(t,x):=\lambda^{-2} \psi(t/\lambda^2,x/\lambda),
\label{eqn:scaling}
\end{align*}
under which the mass and energy corresponding to $\Omega=0$ transform as follows
\begin{subequations}\label{eqn:clas}
\begin{align*}
M_{\Omega=0}(\psi_{\lambda})&=\lambda^{d-4} M_{\Omega=0}(\psi),\\ 
E_{\Omega=0}(\psi_{\lambda})&= \lambda^{d-6} E_{\Omega=0}(\psi),
\end{align*}
\end{subequations}
hence the SN equation is mass-critical for $d=4$ and energy-critical for $d=6$. Although the scaling symmetry is broken in Eq.\ \eqref{eqn:SNHt} by the harmonic term, these critical dimensions demarcate different behaviours of solutions of the SNH equation as well.
 
 Our long-term goal is to understand the dynamics of solutions of the SNH equation in supercritical dimensions. As the first step, here we consider stationary solutions, as they are expected to play the role of attractors  in the dynamics. Stationary solutions are obtained with the ansatz $\psi(t,x)=e^{-i\omega t} f(x)$, where $f(x)$ is a real-valued function and $\omega$ is a real number called frequency. Substituting this ansatz into Eq.\ \eqref{eqn:SNHt} yields
  \begin{align}\label{eqn:SNHnl}
-\Delta f+\Omega^2 |x|^2 f-A_d \left(\int_{\mathbb{R}^d} \frac{|f(y)|^2}{|x-y|^{d-2}}\, dy\right) f = \omega f\,.
\end{align}
  This nonlinear elliptic equation has been thoroughly studied in subcritical dimensions (we refer to \cite{Mor17} for a comprehensive review). In particular, it was proved in \cite{Cao12} for $d<6$ that for each $\omega<d$ there exists a positive, radially symmetric and decreasing to zero solution (such solution will be called a ground state). The proofs in \cite{Cao12} and related works \cite{Fen16, Luo19, Wan08} are based on variational methods. Unfortunately, these methods are not available in supercritical dimensions (technically, the relevant Sobolev embeddings needed to prove existence of critical points of certain functionals cease to be compact). Probably for this reason, to the best of our knowledge,  solutions of Eq.\ \eqref{eqn:SNHnl} for $d>6$ have not been studied in the literature. The goal of this work is to  fill this gap under the assumption of spherical symmetry. For $f=f(r)$, where $r=|x|$, Eq.\ \eqref{eqn:SNHnl} reduces to (hereafter we set $\Omega=1$ by the choice of units)
  \begin{align}\label{eqn:SNHnlrad}
-f''-\frac{d-1}{r} f'+r^2 f&\nonumber\\
-\frac{1}{d-2} \left(\int_{0}^{\infty} \frac{|f(s)|^2}{\max\{r,s\}^{d-2}}\, d y\right) f &= \omega f,
\end{align}
where we have used the Newton formula
\begin{align}\label{eqn:newton}
 \int_{\mathbb{R}^d} \frac{|f(y)|^2}{|x-y|^{d-2}}\, dy=\int_{\mathbb{R}^d} \frac{|f(y)|^2}{\max\{|x|^{d-2},|y|^{d-2}\}}\, dy\nonumber\\
 =\frac{2\pi^{d/2}}{\Gamma(d/2)}\, \int_0^{\infty} \frac{|f(s)|^2 s^{d-1}}{\max\{r^{d-2},s^{d-2}\}}\, ds.
\end{align}
To prove existence, uniqueness and various properties of solutions of Eq.\ \eqref{eqn:SNHnlrad} we shall employ ODE techniques, in particular the shooting method. Similar methods were used for various supercritical nonlinear elliptic equations on  bounded domains \cite{Bud87, Dol07, Guo11, Miy14} and for the supercritical Gross-Pitaevskii equation with the harmonic potential \cite{Bizip, Sel13}.
We remark that in the case of ground states (i.e. positive $f$), the assumption of spherical symmetry does not lead to the loss of generality.  This follows from Theorem 1 in \cite{Bus00} which states  that positive, decaying to zero solutions of semilinear elliptic equations in $\mathbb{R}^d$ must be spherically symmetric, provided they satisfy some additional conditions, which are easy to verify.

The rest of the paper is organized  as follows. In Section \ref{sec:exist} we prove that for every central value $b=f(0)$ there exists a unique ground state with frequency $\omega_0$.  We also show that for each $b$ there exist a sequence $\omega_n$ ($n=1,2,..$) such that the corresponding solutions, called excited states, have exactly $n$ zeros and decay to zero at infinity.  Section \ref{sec:singular} is devoted to singular solutions, i.e.\ solutions which diverge at the origin. We prove the existence of the singular ground state and infinitely many excited states. These results are used to determine the asymptotic behavior of regular stationary states for large values of $b$. Section \ref{sec:omegab} investigates the function $\omega(b)$ for the ground state. We prove that this function is continuous and determine its behaviour for  small and  large values of $b$, observing a qualitatively different behavior in dimensions $7\leq d\leq 15$ and $d\geq 16$.  The paper is concluded with section \ref{sec:summary}, where we summarize the results and discuss open problems that we plan to address in future work. While the paper focuses on supercritical dimensions, along the way we mention relevant results in the critical case $d=6$.

\section{Existence of stationary solutions}\label{sec:exist}
It is routine to show that Eq.\ \eqref{eqn:SNHnlrad} has a one-parameter family of smooth local solutions near the origin
\begin{equation*}\label{icf}
  f(r)=b+\mathcal{O}(r^2),
\end{equation*}
where $b>0$ is a free parameter. For each $b$ we want to find the value(s) of $\omega$ for which the local solution extends to a global smooth solution decaying to zero at infinity. Such solution will be called a ground state if $f(r)$ is positive and an excited state if it has zeros.

Reinstating the potential $v(r)$, we can rewrite Eq.\ \eqref{eqn:SNHnlrad} as the system
\begin{subequations}\label{eqn:SNHlrad}
\begin{align}[left ={ \empheqlbrace}]
 -f''-\frac{d-1}{r} f'+r^2 f+f v &=\omega f,\label{eqn:SNHlrada}\\
v''+\frac{d-1}{r} v' &= f^2.\label{eqn:SNHlradb}
\end{align}
\end{subequations}
It is convenient to remove $\omega$ from Eq.\ \eqref{eqn:SNHlrada} by defining $h(r)=-v(r)+\omega$. Then system \eqref{eqn:SNHlrad} becomes
\begin{subequations}\label{eqn:SNH2}
\begin{align}[left ={ \empheqlbrace}]
f''+\frac{d-1}{r} f'-r^2 f+f h &=0,\label{eqn:SNH2a}\\
h''+\frac{d-1}{r} h'+ f^2&=0.\label{eqn:SNH2b}
\end{align}
\end{subequations}
This system has a two-parameter family of local solutions
\begin{equation}\label{icf2}
  f(r)=b+\mathcal{O}(r^2),\quad h(r)=c+\mathcal{O}(r^2),
\end{equation}
where $c$ is the second free parameter. We will use $c$ as the shooting parameter, i.e. for a given value of the parameter $b$ we will adjust the parameter $c$, so the local solution \eqref{icf2} extends to a global smooth solution for which $(f(r),h(r))\rightarrow (0,h(\infty))$ as $r\rightarrow \infty$. From this we shall recover the frequency as $\omega=h(\infty)$. Note that $h(r)$ is monotonically decreasing as follows immediately from integration of Eq.\ \eqref{eqn:SNH2b}.

One can show that the only possible behaviours of $f(r)$ are that it either diverges to $\pm\infty$ (for a finite or infinite $r$) or converges to zero. To prove this trichotomy, let us assume that the solution exists for all $r$. Since $h(r)$ is a decreasing function, we have $h(r)<h(0)=c$, hence for $r>\sqrt{|c|}$ the term $h(r)-r^2$ is negative, which implies in turn that $f(r)$ cannot have a positive maximum nor a negative minimum. Thus, from some point on $f(r)$ is monotone and therefore there exists a limit $f(\infty)=\lim_{r\rightarrow\infty}f(r)$ (finite or infinite). If $f(\infty)$ is finite, then it must be zero, since otherwise the integral on the right hand side of 
\begin{equation*}
f'(r) = \frac{1}{r^{d-1}} \int_0^\infty [s^2-h(s)]f(s) s^{d-1} ds
\end{equation*}
diverges and L'Hospital's rule gives us $|f'(r)|\to\infty$ as $r\to\infty$, in contradiction with assumed convergence of $f$. This reasoning greatly limits possible behaviours of the solution, telling us that once the solution approaches a positive minimum or negative maximum, it diverges to infinity. Also, from the same line of thought it follows that while $f$ is positive and decreasing (or analogously, negative and increasing), it cannot have an inflection point.

\subsection{Ground states}\label{subsec:ground}
We are going to prove that for every $b>0$ there exists $c$ for which $f(r)$ is positive and  monotonically decays to zero, while $h(r)$ monotonically decays to a constant. We do it in three steps.

\textit{Step 1.} Let us fix $c=0$ and assume that in this case $f$ crosses zero at some $R>0$. Then, multiplying Eq.\ (\ref{eqn:SNH2a}) by $f\, r^{d-1}$ and $f' r^d$, respectively and Eq.\ (\ref{eqn:SNH2b}) by $h\, r^{d-1}$ and $h' r^d$, respectively, and integrating over the interval $[0,R]$ yields four identities
\begin{widetext}
\begin{subequations}\label{eqn:smallc}
\begin{align}
-\int_0^R f'^2 r^{d-1} dr-\int_0^R r^2 f^2 r^{d-1} dr+\int_0^R f^2 h r^{d-1} dr&=0,\label{eqn:smallca}\\
f'(R)^2 R^d+(d-2)\int_0^R f'^2 r^{d-1} dr-\int_0^R f^2 h' r^d dr+(d+2)\int_0^R r^2 f^2 r^{d-1} dr-d \int_0^R f^2 h r^{d-1} dr&=0,\label{eqn:smallcb}\\
h'(R)h(R) R^{d-1}-\int_0^R h'^2 r^{d-1} dr+\int_0^R f^2 h r^{d-1} dr&=0,\label{eqn:smallcc}\\
h'(R)^2 R^d+(d-2)\int_0^R h'^2 r^{d-1} dr+2\int_0^R f^2 h' r^d dr&=0.\label{eqn:smallcd}
\end{align}
\end{subequations}
Taking the combination:  $(d+2)$ $\times$ \eqref{eqn:smallca} + 2 $\times$ \eqref{eqn:smallcb} + $(d-2)$ $\times$ \eqref{eqn:smallcc} + \eqref{eqn:smallcd} yields
\begin{align}
(d-6)\int_0^R f'^2 r^{d-1} dr+(d+2)\int_0^R r^2 f^2 r^{d-1} dr+2f'(R)^2 R^d+(d-2)h(R)h'(R)R^{d-1}+h'(R)^2 R^d&=0.\label{eqn:poh}
\end{align}
\end{widetext}
Since $h(r)$ is decreasing and $h(0)=c=0$, we have $h(R)<0$ and $h'(R)<0$. Hence, for $d\geq 6$ each term in the identity \eqref{eqn:poh} is positive, which gives a contradiction. As a result we see that solution $f(r)$ is positive for $c=0$.

\textit{Step 2.} We now focus on large positive $c$ values and introduce $y=\sqrt{c}r$, $\tilde{f}(y)=f(r)$, and $\tilde{h}(y)=h(r)/c$. In these variables system \eqref{eqn:SNH2} translates to
\begin{align*}[left ={ \empheqlbrace}]
\tilde{f}''+\frac{d-1}{y} \tilde{f'}- \frac{y^2}{c^2} \tilde{f}+ \tilde{f} \tilde{h}&=0,\\
\tilde{h}''+\frac{d-1}{y} \tilde{h'}+\frac{1}{c^2}\tilde{f}^2&=0, 
\end{align*}
with initial conditions
\begin{align*}
\tilde{f}(y)=b+\mathcal{O}(y^2),\qquad \tilde{h}(y)=1+\mathcal{O}(y^2).\label{eqn:boundcinf}
\end{align*}
In the limit $c\to\infty$, the exact solution is
\begin{align*}\label{eqn:cinfsol}
\tilde{f}_{\infty}(y)=b \,\Gamma\left(\frac{d}{2}\right) \frac{2^{\frac{d}{2}-1}}{y^{\frac{d}{2}-1}} J_{\frac{d}{2}-1}(y),\qquad \tilde{h}_\infty(y)=1,
\end{align*}
where $J_q(y)$ is a Bessel function of the first kind. Since $J_{\frac{d}{2}-1}(y)$ is an oscillating function and $\tilde{f}(y)$ tends uniformly to $\tilde{f}_{\infty}(y)$ on every compact interval as $c\to\infty$, it follows that if $c$ is sufficiently large, then the solution $f(r)$ crosses zero arbitrarily many times.

\textit{Step 3.} Let us define
\begin{align*}
I_0=\{&c\geq 0\, | \, \exists\, r_0>0 : f(r_0)=0\mbox{ while }\\
&f(r)>0, f'(r)<0 \mbox{ for } r\in(0,r_0)\}.
\end{align*}
It follows from Step 2 that $I_0$ is nonempty, while Step 1 gives us a lower bound for this set, since $0\notin I_0$. Hence, we may define $c_0=\inf I_0\geq0$. Let  $f_0(r)$ be  the solution with $c=c_0$. We first  show that $f_0(r)$ cannot cross zero. Assume otherwise that $f(r_0)=0$ for some $r_0$. Then, by the continuous dependence of solutions on the initial condition solutions with $c$ close to $c_0$ also cross zero near $r_0$ (the potentially problematic situation when $f'(r_0)=f(r_0)=0$ is excluded because then $f(r)\equiv 0$). As there cannot be an inflection point in $(0,r_0)$, the function stays decreasing for these $c$ and we have a contradiction with $c_0$ being an infimum of $I_0$. Now assume that $f_0(r)\to\infty$ as $r\to\infty$. Since $f_0(r)$ is initially decreasing and positive, there must exist a single positive minimum $r_1$. From the continuous dependence of solutions on the initial conditions it follows that there exists a small neighbourhood of $c_0$ with no elements in $I_0$. This again contradicts that  $c_0$ is an infimum of $I_0$. Due to the trichotomy we conclude that $f_0(r)>0$, $f'(r)<0$, and $\lim_{r\to\infty}f_0(r)= 0$. The latter implies that the nonlinear term in Eq.\ \eqref{eqn:SNHnlrad} is negligible for large $r$ and therefore $f(r)$ decays exponentially for large $r$. This and integration of Eq.\ \eqref{eqn:SNH2b} implies that $\lim_{r\rightarrow\infty} h(r)$ is finite.

Having this result and taking $\omega=h(\infty)$, we recover the ground state solution of Eq.\ \eqref{eqn:SNHlrad} with $v(r)=\omega-h(r)$, where $v$ vanishes in infinity as one could expect from a potential. This gives us one-to-one correspondence between the formulations  \eqref{eqn:SNHlrad} and \eqref{eqn:SNH2}.
Finally, let us point out that since Step 1 holds also for $d=6$  and other results used in the proof do not depend on dimension $d$, this theorem holds also in the critical case $d=6$.

\subsection{Uniqueness of ground states}
Ground states found in the previous subsection are unique in the sense that for any value of $b>0$ there exists exactly one value of $c$ as described. To show it, we use the argument coming from a proof of Proposition 1.1 in \cite{Gal11}. Let us assume that system (\ref{eqn:SNH2}) has two positive solutions, $f_1$ and $f_2$, such that
\begin{align*}
f_i(0)=b,\qquad h_i(0)=c_i, \qquad f_i'(0)=h_i'(0)=0,
\end{align*}
where $i\in\{1,2\}$. Since $f_i>0$, we may define $\rho(r)=f_1(r)/f_2(r)$. Then $\rho(0)=1$ and $\rho'(0)=0$. It is convenient to introduce $\mu(r)= r^{d-1} f_2(r)^2 \rho'(r)$, then one may show that
\begin{align*}
\mu'(r)=r^{d-1} f_2(r)^2\rho(r)\left( h_2(r)-h_1(r)\right). \label{eqn:rhoprime}
\end{align*}
Without the loss of generality we may assume that $c_1>c_2$. As $\mu'(0)=0$, then there exists $r_0>0$ such that $\mu'(r)<0$ for $r\in (0,r_0)$. Since $\mu(0)=0$, also $\mu(r)<0$ in $r\in (0,r_0)$, hence $\rho$ is initially decreasing.

Let us now look into $h_i$. If we define $\delta=h_2-h_1$, we have $\delta(0)=c_2-c_1<0$ and $\delta'(0)=0$. This function satisfies the equation
\begin{align*}
\left(r^{d-1} \delta'\right)'+r^{d-1} f_2^2(1-\rho^2)=0.
\end{align*}
It means that as long as $\rho<1$, e.g.\ for $r\in(0,r_0)$, $\delta$ is decreasing.
Hence, if for some interval beginning in zero it holds $\rho<1$, then also $h_1>h_2$ in it.

Let us now assume that at some point $\rho'(r)>0$ and define $r_1 := \inf\{r>0\, |\, \rho'(r)=0\}$. Then for $r\in(0,r_1)$ we have $\rho(r)<1$, $h_1(r)>h_2(r)$, and $\mu'(r)<0$. It contradicts the fact that $\mu(r)<0$ in this interval, while $\mu(r_1)=0$. Hence, $\rho'<0$ everywhere and $\mu$ is a decreasing function.

From the monotonicity of $\mu$ for $r>1$ we have  $r^{d-1} f_2(r)^2 \rho'(r)<f_2(1)^2 \rho'(1)<0$, so for such $r$:
\begin{align*}
\rho'(r)<\frac{f_2(1)^2 \rho'(1)}{r^{d-1} f_2(r)^2}<0
\end{align*}
Since $\rho>0$ and $\rho(1)<1$, we have
\begin{align*}
-1&< \lim_{r\to\infty} \rho(r) - \rho(1) = \int_{1}^{\infty} \rho'(r)\, dr\nonumber\\
&< f_2(1)^2 \rho'(1) \int_1^\infty \frac{dr}{r^{d-1} f_2(r)^2}<0.
\end{align*}
Hence, the right hand side integral is finite. Since $f_2$ decays exponentially, there is
\begin{align*}
\infty&=\int_1^\infty dr=\int_1^\infty  r^{d-1} f_2(r)^2 \frac{1}{r^{d-1} f_2(r)^2}\, dr\nonumber\\
&\leq \left(\int_0^\infty r^{d-1} f_2(r)^2\, dr\right)^{1/2} \left(\int_0^\infty \frac{dr}{r^{d-1} f_2(r)^2}\right)^{1/2}<\infty.
\end{align*}
This contradiction means that $c_1=c_2$ and the solution is unique.

This result does not depend on dimension $d$, hence it works also in the critical case. Together with one-to-one correspondence between $c$ and $\omega$ it lets us define a function $\omega(b)$. We will investigate its properties in Section \ref{sec:omegab}.

\subsection{Excited states}
The proof of existence of ground states in critical and supercritical dimensions can be generalised to give us also spherically symmetric excited states. More precisely, we will see that for every natural number $n$ there exists a value of $c$ such that the solution $f$ of Eq.\ (\ref{eqn:SNH2}) crosses zero exactly $n$ times. We begin by defining a set similar to $I_0$, this time with at least two zeroes separated by a minimum:
\begin{align*}
I_1=\{&c\geq 0\, | \, \exists\, r_0, r_1,r_2>0 :\,r_0<r_1<r_2,\,f'(r_1)=0,\\
&f(r_0)=f(r_2)=0,\, f'(r)<0 \mbox{ for } r\in(0,r_1), \\
&\mbox{ and } f'(r)>0 \mbox{ for } r\in(r_1,r_2)\}.
\end{align*}
From the behaviour of solutions for large $c$ we know that $I_1$ is nonempty. The definition implies $I_1\subset I_0$, so $c_1 := \inf I_1 \geq c_0$. Let $f_1$ be a solution for this $c=c_1$.

The solution cannot be tangent to zero line at any moment, so as $c$ changes, the only way for new zeroes to appear is to come from infinity. From the proof of trichotomy we know that if solution crosses zero at $r>\sqrt{c}$, it must blow up to infinity then, so new zeroes appear individually. Knowing this and using the fact that there can be no inflection point while solution is negative and increasing, we conclude that $f_1$ crosses zero at least once. Then we can repeat the reasoning from the previous proof, i.e. by using the continuous dependence of the solution on the initial conditions we see that $f_1$ cannot cross zero for the second time, so it either monotonically converges to some nonpositive value or it bends down at some point $r_2$ such that $f(r_2)<0$ and $f'(r_2)=0$. The second option contradicts $c_1$ being the infimum, so the trichotomy gives $\lim_{r\to\infty}f_1(r)= 0$. Defining $I_n$ for higher values of $n$ in a similar manner, one can repeat this deduction.

Even though the numerical results suggest that spherically symmetric excited states are unique, one cannot employ the method used for ground states to show it formally. In fact, uniqueness of excited states is in general a rather complicated problem, unsolved even in case of simpler systems than SNH (cf.\ \cite{Has11}).

Let us point out that the reasoning used in both proofs of existence is based on some rather general presumptions. The main roles were played by three observations: trichotomy of $\lim_{r\to\infty} f(r)$, positivity of the solution for $c=0$, and existence of oscillations in the limit $c\to\infty$. It suggests that this proof may be easily modified to show the existence of the ladder of solutions in case of some other nonlinear problems. The examples of such problems are singular solution of SNH, considered in the next section, and Gross-Pitaevskii equation.

\section{Singular solutions}\label{sec:singular}
To get singular solutions of our system, i.e.\ solutions of Eq.\ (\ref{eqn:SNH2}) such that $\lim_{r\to 0} f(r) = \infty$, we begin with an introduction of rescaled variables $\rho=\sqrt{b}r$, $F(\rho)=b^{-1} f(r)$, and $H(\rho)=b^{-1} h(r)$. Then
\begin{align*}[left ={ \empheqlbrace}]
F'' + \frac{d-1}{\rho} F' - b^{-2} \rho^2 F + F H&= 0,\\
H''+\frac{d-1}{\rho} H' +F^2&= 0.
\end{align*}
For a fixed value of $\rho$ we may perform a limit $b\to\infty$ obtaining a system of two equations possessing a synchronised solution $F=H$, satisfying a quadratic Lane-Emden equation in $d$ dimensions:
\begin{align}\label{eqn:binf1}
F'' + \frac{d-1}{\rho} F' + F^2&= 0.
\end{align}
It has a singular solution $F(\rho)=\frac{2(d-4)}{\rho^2}$, that can be converted back to $f_\infty(r)=\frac{2(d-4)}{r^2}$, suggesting introduction of new functions such that
\begin{align}\label{eqn:fftilde}
f(r)=\frac{2(d-4)}{r^2} \tilde{f}(r),\qquad h(r)=\frac{2(d-4)}{r^2} \tilde{h}(r),
\end{align}
satisfying
\begin{subequations}\label{eqn:ftilde}
\begin{align}[left ={ \empheqlbrace}]
\tilde{f}'' + \frac{d-5}{r}\tilde{f}' +\frac{2(d-4)}{r^2}(\tilde{f}\tilde{h}-\tilde{f})-r^2 \tilde{f}&= 0,\label{eqn:ftildea}\\
\tilde{h}''+\frac{d-5}{r} \tilde{h}'+ \frac{2(d-4)}{r^2}(\tilde{f}^2-\tilde{h})&= 0\label{eqn:ftildeb}.
\end{align}
\end{subequations}
This procedure lets us to factor out the singular behaviour. In the next subsection we investigate behaviour of $\tilde{f}$ and $\tilde{h}$ near zero.

\subsection{Asymptotic behaviour near zero}
Let us transform system (\ref{eqn:ftilde}) by an introduction of $t=\ln r$ (so we now focus on $t\to -\infty$), $\eta=\tilde{f}-1$ and $\xi=\tilde{h}-1$. We get
\begin{subequations}\label{eqn:etaxi}
\begin{align}[left ={ \empheqlbrace}]
\ddot{\eta} + (d-6)\dot{\eta} +2(d-4)\xi= e^{4t}(1+\eta) -2(d-4)\eta\xi,\label{eqn:etaxia}\\
\ddot{\xi} + (d-6)\dot{\xi} +2(d-4)(2\eta-\xi) = -2(d-4)\eta^2.\label{eqn:etaxib}
\end{align}
\end{subequations}
The linear system on the left hand side has four eigenvalues:
\begin{align*}
\lambda_1^\pm =& \frac{-d+6\pm \sqrt{d^2-20d+68}}{2},\\
\lambda_2^\pm =& \frac{-d+6\pm\sqrt{d^2+4d-28}}{2}.
\end{align*}
In supercritical cases $(d\geq 7)$, the real parts of $\lambda_1^\pm$ are always negative. For $d<2 (5 + 2 \sqrt{2})\approx 15.66$ they have a non-zero imaginary part, while for larger $d$ they are real numbers. On the other hand, for $d\geq 7$ eigenvalues $\lambda_2^\pm$ are a pair of real numbers, one negative and one positive. In the following, we will be often using $\lambda_2^+$ so for the future convenience let us denote it as $\lambda$ and point out that for $d\geq 7$ there is $3\leq \lambda<4$. The considered linear system is hyperbolic and the stable manifold theorem \cite{Cod72} tells us that there exists a one-dimensional unstable manifold with solutions behaving like $e^{\lambda t}$ as $t\to -\infty$. The existence of this manifold is a feature significantly distinguishing our system from the Gross-Pitaevskii equation and similar nonlinear Schr\"{o}dinger equations \cite{Bizip, Mer91, Sel13}. As we will see, its parametrisation will serve us as a suitable shooting parameter. We may obtain it using reasoning similar to the proof of Lemma 3.1.\ in Ref.\ \cite{Mer91}.

For clarity of the presentation let us focus on the case $d\geq 16$ (if $7\leq d \leq 15$, the analysis is analogous with some minor changes discussed in the end of this subsection). It is convenient to introduce
\begin{align*}
x(t)&=e^{4t}(1+\eta(t))-2(d-4)\eta(t)\xi(t),\\
y(t)&=-2(d-4)\eta(t)^2,
\end{align*}
and also define $\alpha_1=\frac12 \sqrt{|d^2-20d+68|}$, $\alpha_2= \frac12 \sqrt{d^2+4d-28}$, and $\beta=-\frac{d}{2}+3$ (so $\lambda=\beta+\alpha_2$). Then using the method of variation of parameters, one can write the solutions of system (\ref{eqn:etaxi}) implicitly as
\begin{widetext}
\begin{subequations}\label{eqn:etaxisol}
\begin{align}
\eta(t)=&-c e^{\lambda t}+\frac{1}{3\alpha_1} \int_{-\infty}^t [2x(s)+y(s)]e^{\beta(t-s)} \sinh\alpha_1(t-s)ds+\frac{1}{3\alpha_2} \int_{-\infty}^t [x(s)-y(s)]e^{\beta(t-s)} \sinh\alpha_2(t-s)ds,\label{eqn:etaxisola}\\
\xi(t)=&2c e^{\lambda t}+\frac{1}{3\alpha_1} \int_{-\infty}^t [2x(s)+y(s)]e^{\beta(t-s)} \sinh\alpha_1(t-s)ds-\frac{2}{3\alpha_2} \int_{-\infty}^t [x(s)-y(s)]e^{\beta(t-s)} \sinh\alpha_2(t-s)ds,\label{eqn:etaxisolb}
\end{align}
\end{subequations}
\end{widetext}
where $c$ is some parameter. We discarded all terms coming from the homogenous part, other then ones proportional to $e^{\lambda t}$, as we are interested in solutions decaying in $-\infty$.

Also from the decay of $\eta$ and $\xi$ we know that for every $\varepsilon>0$ one may find such $T$ that for all $t<T$ the following boundaries exist:
\begin{align*}
|2x(t)+y(t)|&=2\left|e^{4t}(1+\eta(t))+(d-4)\eta(t)[\eta(t)+2\xi(t)]\right|\nonumber\\
&\leq 4 e^{4t}+\varepsilon |\eta(t)|,\\
|x(t)-y(t)|&=\left|e^{4t}(1+\eta(t))+2(d-4)\eta(t)[\eta(t)-\xi(t)]\right|\nonumber\\
&\leq 2 e^{4t}+\varepsilon |\eta(t)|.
\end{align*}
Plugging them into Eq.\ (\ref{eqn:etaxisol}) and evaluating necessary integrals lets us to constrain one of the solutions as
\begin{align}\label{eqn:etabound}
|\eta(t)|\leq |c| e^{\lambda t} + A_1 e^{4 t}+\varepsilon A_2 e^{\lambda t} \int_{-\infty}^t e^{-\lambda s}|\eta(s)|ds,
\end{align}
with $A_1$ and $A_2$ being some positive constants. In a similar way one gets a constraint on $|\xi(t)|$. We may multiply both of these bounds by $e^{-\lambda t}$ and then use the integral Gr\"{o}nwall's inequality to get
\begin{align*}
|\eta(t)| \leq |c| e^{\lambda t} + B_1 e^{4 t},\qquad |\xi(t)| \leq |2c| e^{\lambda t} + B_2 e^{4 t},
\end{align*}
for sufficiently small $t$ values, where $B_1$ and $B_2$ also denote positive constants. Then definitions of $x$ and $y$ result, for sufficiently small $t$, in:
\begin{align*}
x(t)= e^{4t} + \mathcal{O}(e^{2\lambda t}),\qquad y(t)= \mathcal{O}(e^{2\lambda t}),
\end{align*}
where we used the fact that for $d\geq 7$ it holds $\lambda<4<2\lambda$. In the end, plugging these results into Eq.\ (\ref{eqn:etaxisol}) and integrating yields 
\begin{align}\label{eqn:etaxias}
\eta(t)=-c e^{\lambda t} + \mathcal{O} (e^{4t}),\qquad \xi(t)=2ce^{\lambda t} + \mathcal{O} (e^{4 t})
\end{align}
or, returning to the previous variables,
\begin{align}\label{eqn:asymorig}
\tilde{f}(r)=1-c r^\lambda + \mathcal{O} (r^4),\qquad \tilde{h}(r)=1+2c r^\lambda + \mathcal{O} (r^4).
\end{align}

For $7\leq d \leq 15$ the proof is almost identical, with the main difference being negativity of the quadratic expression present in $\alpha_1$. To take it into account one needs to change hyperbolic sines present in Eq.\ (\ref{eqn:etaxisol}) into sines. Since sine is a bounded function, one can obtain Eq. (\ref{eqn:etabound}) and the latter bounds as well.

\subsection{Existence of singular solutions}
The just found parametrisation of the unstable manifold will play a role of a shooting parameter in our proof of the fact that in $d\geq 7$ for each nonnegative integer $n$ there exists a solution of system (\ref{eqn:ftilde}) with behaviour near $-\infty$ given by Eqs.\ (\ref{eqn:etaxias}) and with function $\tilde{f}$ crossing zero exactly $n$ times before decaying to zero in infinty. After returning to the initial variables (\ref{eqn:fftilde}) such solutions give the whole ladder of singular solutions of system (\ref{eqn:SNH2}), beginning with the ground state. We perform this proof in four steps, where steps 1--3 follow the steps of the proof of existence of regular ground states, while the last step covers excited states.

\textit{Step 1.} We fix $c<0$. Then in some neighbourhood of zero solutions of system (\ref{eqn:ftilde}) satisfy $\tilde{f}'>0$ and $\tilde{h}'<0$, hence it holds $\tilde{f}>1$ and $\tilde{h}<1$ as long as signs of the derivatives do not change. From Eq.\ (\ref{eqn:ftildea}) it is clear that $\tilde{f}'$ cannot change its sign if $\tilde{h}'$ did not change its earlier. Analogous conclusion can be made for $\tilde{h}'$ with Eq.\ (\ref{eqn:ftildeb}). As a result we see that for any $c<0$, function $f$ is a strictly positive, increasing solution.

\textit{Step 2.} Following proof from Section \ref{subsec:ground}., we now look for solutions with large positive $c$. Thus we introduce a new independent variable $s=\ln( c^{1/\lambda} r)$, so that in a limit $c\to\infty$ Eq.\ (\ref{eqn:ftilde}) becomes an autonomous system:
\begin{subequations}\label{eqn:ftilde2}
\begin{align}[left ={ \empheqlbrace}]
\ddot{\tilde{f}} + (d-6)\dot{\tilde{f}} +2(d-4)(\tilde{f}\tilde{h}-\tilde{f})&= 0,\label{eqn:ftilde2a}\\
\ddot{\tilde{h}}+(d-6) \dot{\tilde{h}}+2(d-4)(\tilde{f}^2-\tilde{h})&= 0,\label{eqn:ftilde2b}
\end{align}
\end{subequations}
with solutions behaving near $-\infty$ as
\begin{align*}
\tilde{f}(s)=1- e^{\lambda s} + \mathcal{O} (e^{4s}),\qquad \tilde{h}(s)=1+2e^{\lambda s} -\mathcal{O} (e^{4 s}).
\end{align*}
We define an energy $E$ of this system as
\begin{align*}
E= \dot{\tilde{f}}^2+\frac{1}{2}\dot{\tilde{h}}^2-2(d-4)\tilde{f}^2-(d-4)\tilde{h}^2+2(d-4)\tilde{h}\tilde{f}^2.
\end{align*}
One can easily show that such quantity is decreasing with $s$ and $\lim_{s\to -\infty} E(s)=-(d-4)$. Hence, for any $s$ we have $-(d-4)>(d-4)(-2\tilde{f}^2-\tilde{h}^2+2\tilde{h}\tilde{f}^2)$ and so $(\tilde{h}-1)(\tilde{h}+1-2\tilde{f}^2)>0$. Initially $\tilde{h}$ is increasing, starting from 1, so in some interval $(-\infty,s_0)$ it holds $\tilde{h}+1>2\tilde{f}^2$. It implies $\tilde{f}^2-\tilde{h}<\frac12 (1-\tilde{h})$ and results in negativity of the last term of Eq.\ (\ref{eqn:ftilde2b}) in such interval. It means that $\tilde{h}$ cannot have local maximum and in consequence is an increasing function. In conclusion, Eq.\ (\ref{eqn:ftilde2a}) is a damped linear oscillator with increasing frequency and $\tilde{f}$ oscillates with decreasing amplitude.

\textit{Step 3.} The asymptotic behaviour (\ref{eqn:asymorig}) translates to $h$ as
\begin{align*}
h'(r) = -\frac{4(d-4)}{r^3}+4c (d-4)(2-\lambda)r^{\lambda-3}+\mathcal{O}(r).
\end{align*}
Since $d>6$ and $3\leq \lambda < 4$, function $h' r^{d-1}$ tends to zero as $r\to 0$. Then we have
\begin{align*}
h'(r) = -\frac{1}{r^{d-1}} \int_0^r \rho^{d-1} f(\rho)^2 d\rho
\end{align*}
and so $h$ is decreasing also in singular case. As a result, $\tilde{h}/r^2$ is bounded from above and the reasoning similar to the one that gave us a trichotomy in the regular case also works here (giving us the secondary results connected to extrema and inflection points as well). These results together with outcomes of Step 1 and Step 2 let us repeat the proof of existence of a ground state in the exact same way.

\textit{Step 4.}
Since solution $\tilde{f}$ cannot be tangent to the zero line (or otherwise it is zero), as $c$ variates new zeros may only appear by coming from infinity. This observation, together with results mentioned in Step 3 (after a positive minimum, or negative maximum, $\tilde{f}$ blows up to infinity; there are no positive decreasing, or negative increasing, inflection points) is all we need to recreate the proof of existence of excited states.

We can also easily show uniqueness of a just found ground states, i.e.\ uniqueness of a value of $c$ for which the solution $\tilde{f}$ is positive and decays to zero in infinity. To prove this fact, it suffices to repeat the reasoning used in a case of regular solutions, with just a change of appropriate exponents from $d-1$ to $d-5$. Here the assumption $c_1>c_2$ also translates to existence of an interval $(0,r_0)$, where $\rho<1$ and $\delta<0$. The second observation comes from considerations of the equation
\begin{align*}
\left(r^{d-5} \delta'\right)'=2(d-4)r^{d-7}[\tilde{f}_2^2(\rho^2-1)+\delta].
\end{align*}
Analogously, we get negativity of $\rho'$ and monotonicity of $\mu$. In the limit $r \to \infty$ we have $\tilde{f}\to 0$ and the third term in Eq.\ (\ref{eqn:ftildea}) is negligible giving us an exponential decay and convergence of appropriate integrals. In effect, we may repeat the last part of the proof obtaining contradiction and in conclusion uniqueness of $c$.

\begin{figure}
\centering
\includegraphics[width=0.45\textwidth]{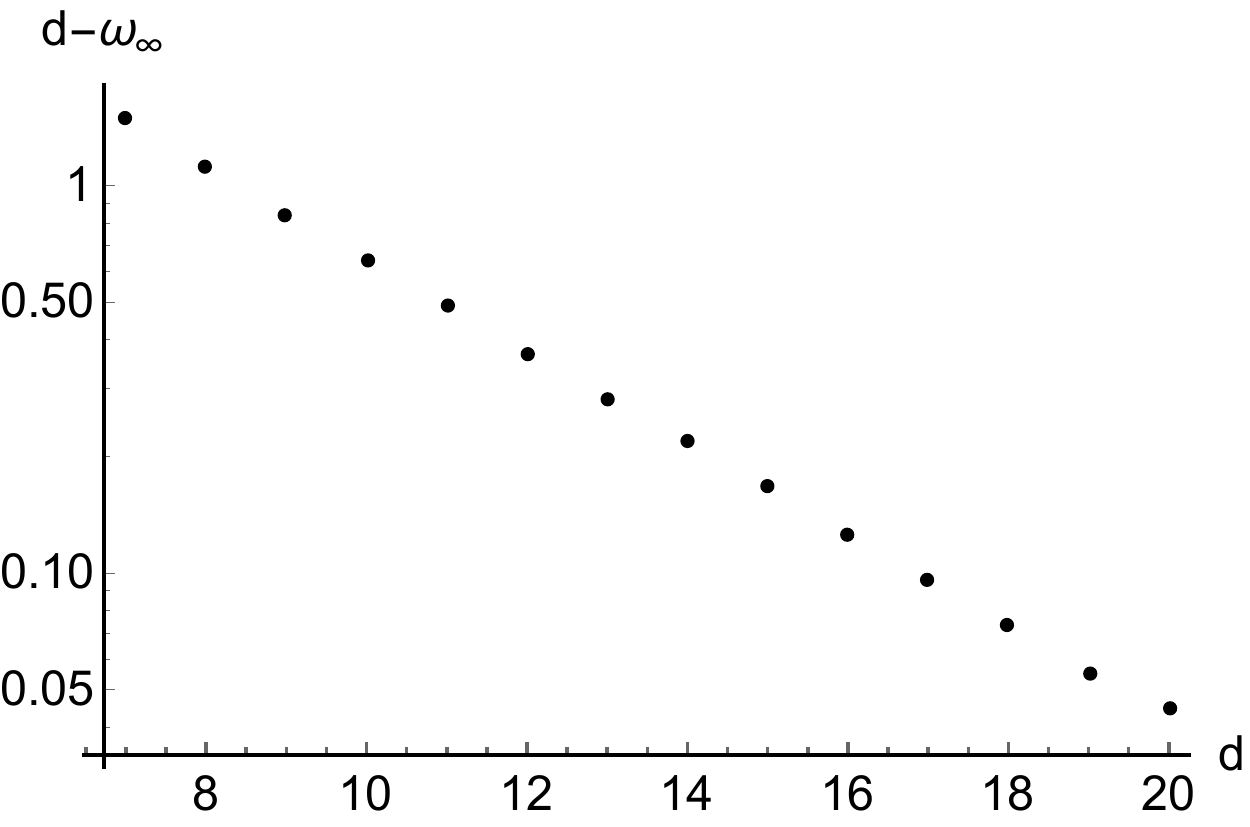}
\caption{Dependence of a difference $d-\omega_\infty$ on dimension $d$. Up to numerical errors, there exists an empiric relation $d-\omega_\infty(d)=A\, e^{-\gamma d}$, where $A\approx 9.64$ and $\gamma\approx 0.271$.}
\label{fig:omegainfty}
\end{figure}

If we rewrite Eq.\ (\ref{eqn:ftildeb}) using variable $t=\ln r$, we get
\begin{equation*}
\ddot{\tilde{h}}+(d-6)\dot{\tilde{h}}+2(d-4)(\tilde{f}^2-\tilde{h})=0.
\end{equation*}
For every interval $(-\infty,t_0)$ one can choose sufficiently large $c$, that $h$ is increasing there. Since $\tilde{f}$ converges to zero we can pick such large values of $t$ and $c$ that this equation can be approximated there by a linear equation $\ddot{\tilde{h}}+(d-6)\dot{\tilde{h}}-2(d-4)\tilde{h}=0$. The dominant behaviour of its solution for large $t$ is $e^{2t}$, which translates to $\tilde{h}\sim r^2$ in original independent variable. It means that the function $h$ converges to some fixed value $\omega_\infty:=\lim_{r\to\infty} h(r)$. If we shift function $h$ by this value, as in the regular case, we obtain a system of singular solutions of system (\ref{eqn:SNHlrad}) with a potential vanishing in infinity and a frequency equal to $\omega_\infty$. Table \ref{tab:omegainfty} gives values of $\omega_\infty$ (for ground states) obtained this way and Figure \ref{fig:omegainfty} plots them.
\begin{table}
\centering
\begin{tabular}{c|ccccccc}
$d$ & 7 & 8 & 9 & 10 & 11 & 12 & 13\\
$\omega_\infty$ & 5.504 & 6.885 & 8.161 & 9.363 & 10.515 & 11.623 & 12.717\\ \hline
$d$ & 14 & 15 & 16 & 17 & 18 & 19 & 20\\
$\omega_\infty$ & 13.783 & 14.834 & 15.873 & 16.903 & 17.926 & 18.945 & 19.955
\end{tabular}
\caption{Values of a frequency $\omega_\infty$ of a singular ground state in various supercritical dimension $d$.
}\label{tab:omegainfty}
\end{table}

Analysis presented in this Section works only for supercritical dimensions of SNH, since in critical dimension the considered linear system loses its hyperbolicity and our reasoning ceases to work.

\section{Behaviour of function $\omega(b)$}\label{sec:omegab}
\begin{figure}
\centering
\includegraphics[width=0.45\textwidth]{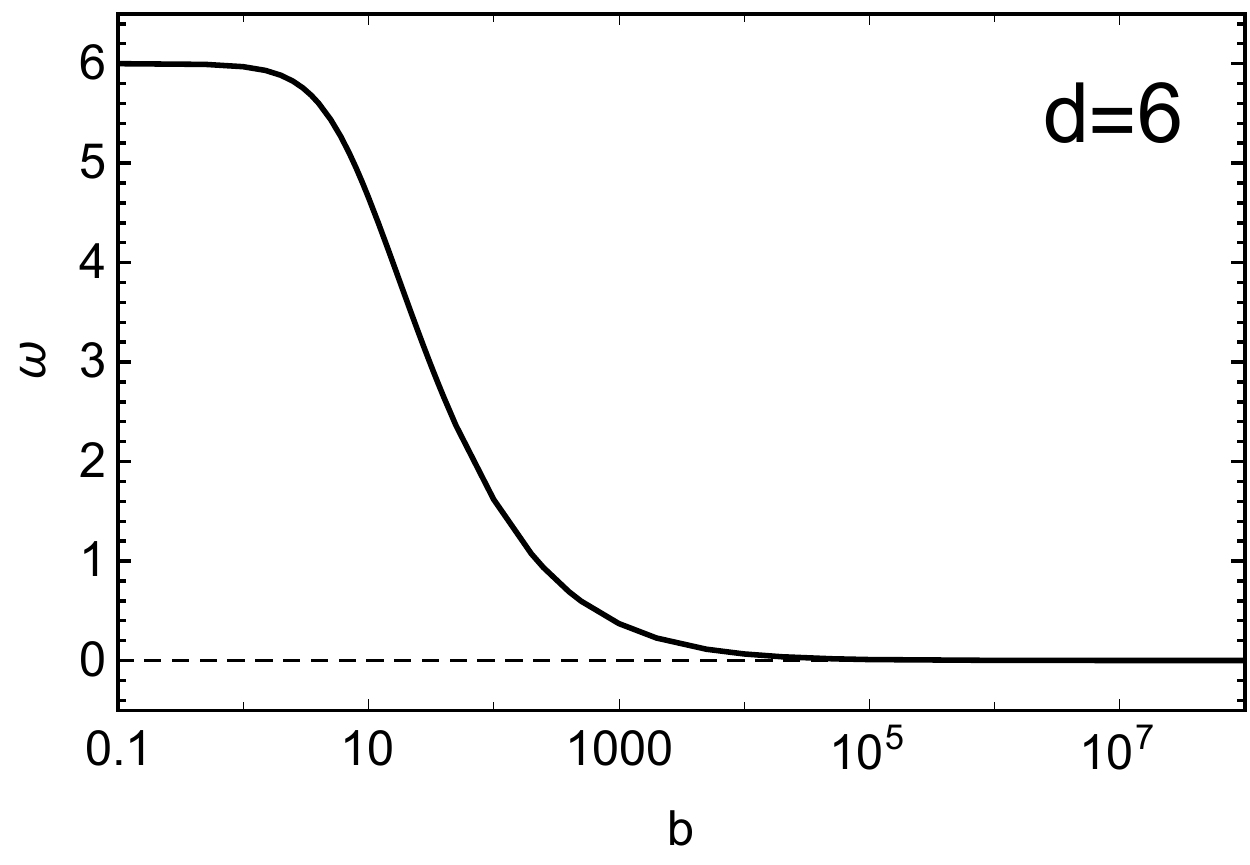}
\includegraphics[width=0.45\textwidth]{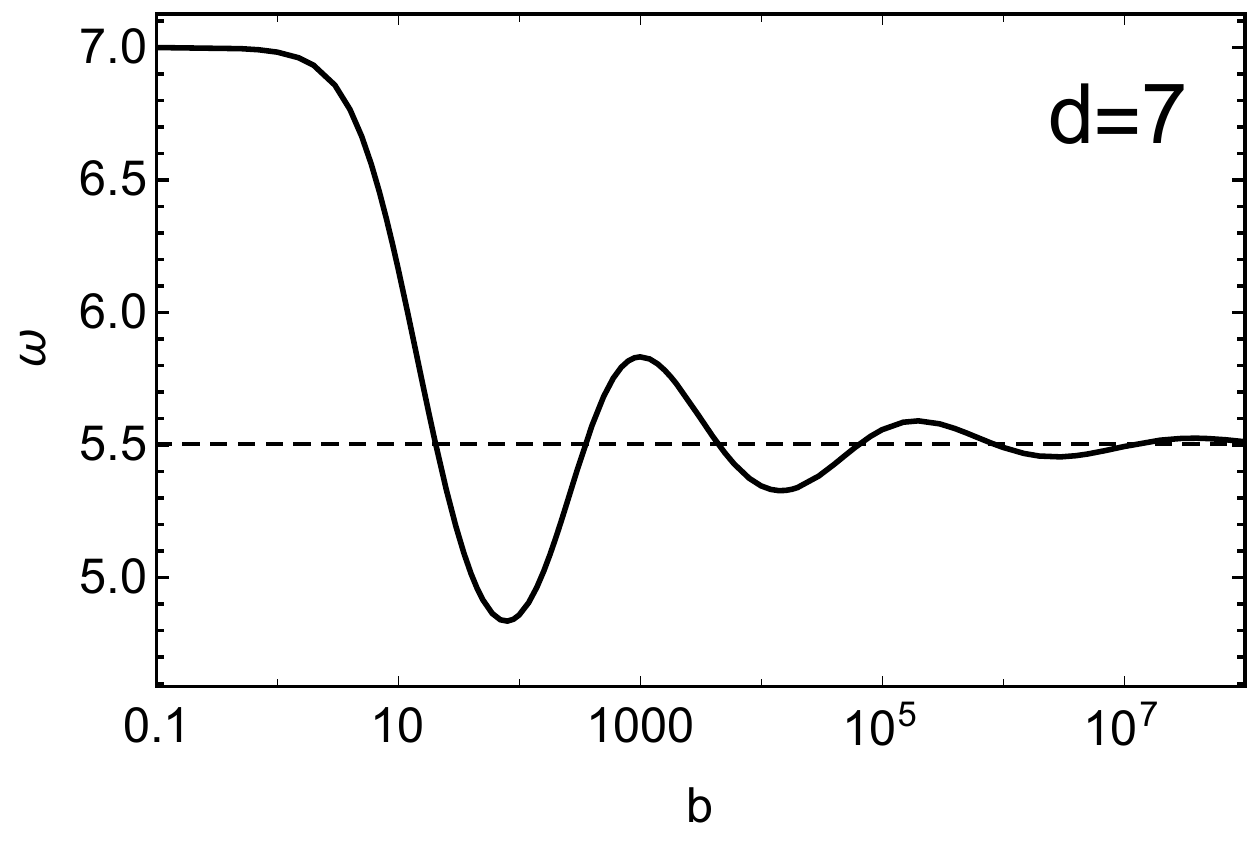}\\
\includegraphics[width=0.45\textwidth]{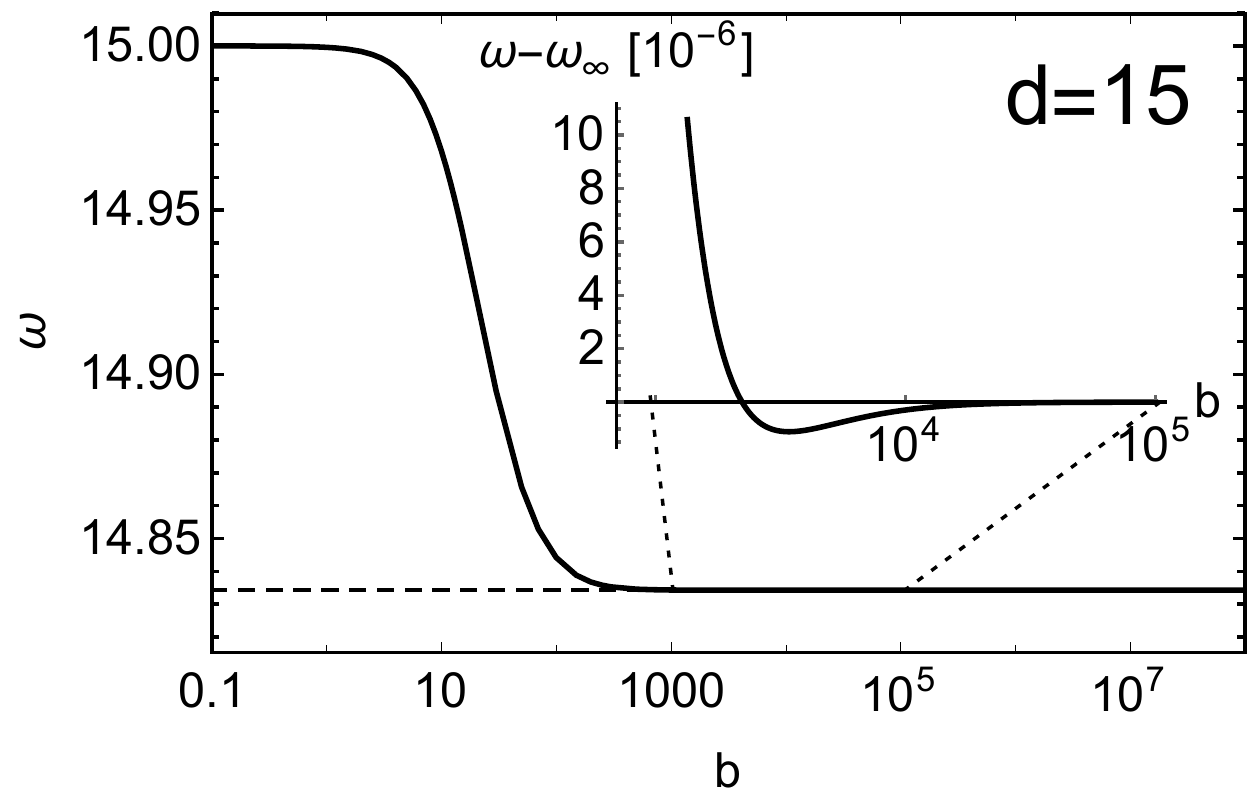}
\includegraphics[width=0.45\textwidth]{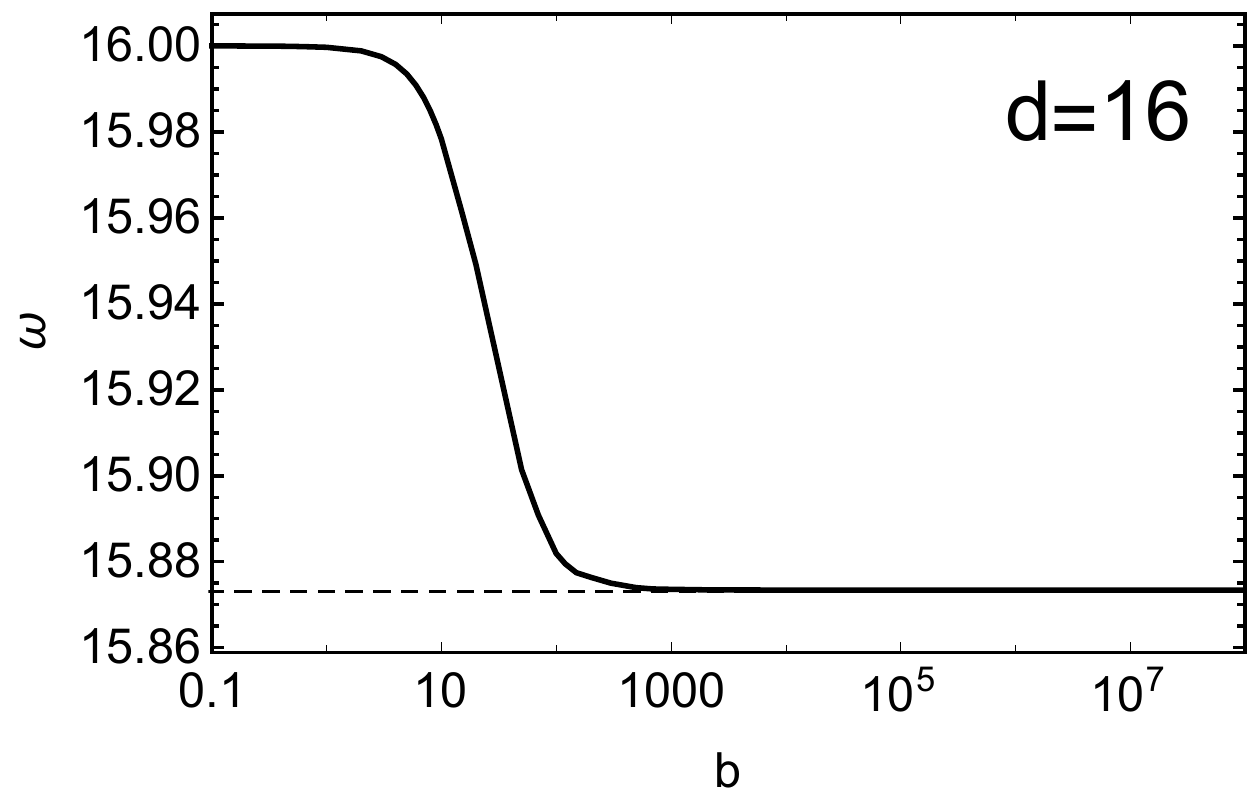}
\caption{Plots of relations between $b$ and $\omega$ for ground state of SNH in critical (top) and supercritical (remaining plots) cases. For $d=15$ the oscillations have so small amplitude that in order to resolve them, the inset of the plot needed to be zoomed in.}
\label{fig:plotd}
\end{figure}

Knowing that for every $b>0$ there exists a unique frequency of a ground state $\omega$, we may define a function $\omega(b)$. It is defined for each dimension $d$ separately and some exemplary plots for various $d$ (including critical and supercritical cases) are showed in Fig.\ \ref{fig:plotd}. In the critical dimension it is a decreasing function approaching zero in infinity. For $d=7$ it is oscillating around $\omega_\infty$, a frequency of the singular ground state, with decreasing amplitude, meaning that there exists an infinite number of ground states with frequency $\omega_\infty$. As $d$ gets bigger, these oscillations become smaller and finally at $d=16$ they completely vanish restoring monotonicity of $\omega(b)$. We now aim to analytically explain some of these behaviours.

\subsection{Allowed range}\label{sec:omegarange}
Let us denote by $e_0$ the function
\begin{align*}\label{eqn:e0}
e_0(r)=\sqrt{\frac{2}{\Gamma(d/2)}}e^{-r^2/2}.
\end{align*}
Then it is a normalized ground state of a quantum linear oscillator $-\Delta e_0+r^2 e_0=d\, e_0$. We have
\begin{align*}
0&=(e_0,\Delta f-r^2 f+\omega f-fv)\nonumber\\
&=(\Delta e_0-r^2 e_0,f)+\omega(e_0,f)-(e_0,f\, v)\nonumber\\
&=(\omega-d)(e_0,f)-(e_0,f\, v),
\end{align*}
where $(\cdot,-)$ is a standard scalar product: $(f,g)=\int_0^\infty f(r) g(r) r^{d-1} dr$. The norm induced by this product will be denoted by $\Vert \cdot \Vert$. Since $e_0$ and $f$ are positive functions, while $v$ is negative, we get $\omega>d$. This simple result gives us an upper bound on $\omega$ value.

The lower bound comes from the Pohozaev-type identities. In a manner similar to Step 1 of the proof of existence of ground states, we multiply Eq.\ (\ref{eqn:SNHlrada}) by $f(r)r^{d-1}$. We may perform integration by parts, with boundary terms vanishing due to the fast decay of $f$, and get
\begin{equation}\label{eqn:Poh1}
-\Vert f' \Vert^2 - \Vert r f \Vert^2 + \omega \Vert f \Vert^2- \int_0^\infty f(r)^2 v(r) r^{d-1} \, dr =0.
\end{equation}
Analogously, multiplying by $f'(r)r^{d}$ and integrating gives
\begin{align}\label{eqn:Poh2}
\frac{d-2}{2} \Vert f' \Vert^2 +\frac{d+2}{2} \Vert r f \Vert^2 +\frac{d}{2} \int_0^\infty f(r)^2 v(r) r^{d-1}  \, dr\nonumber\\
 +\frac{1}{2} \int_0^\infty f(r)^2 v'(r) r^d  \, dr - \omega d \Vert f \Vert^2_2=0.
\end{align}
We perform in an almost identical manner with Eq.\ (\ref{eqn:SNHlradb}), i.e.\ we multiply by $v(r) r^{d-1}$ or $v'(r) r^d$ and integrate, obtaining:
\begin{align}
\Vert v' \Vert^2+ \int_0^\infty f(r)^2 v(r) r^{d-1} \, dr &=0,\label{eqn:Poh3} \\
\frac{d-2}{2}\Vert v' \Vert^2- \int_0^\infty f(r)^2 v'(r) r^d \, dr &=0.\label{eqn:Poh4}
\end{align}
Asymptotic behaviour of the solutions assures convergence of all these integrals. Using Eqs.\ (\ref{eqn:Poh1}) -- (\ref{eqn:Poh4}) to eliminate terms including $\Vert v' \Vert^2$,  $ \int_0^\infty f(r)^2 v(r) r^{d-1}  \, dr$, and $ \int_0^\infty f(r)^2 v'(r) r^d \, dr$ gives us a so-called Pohozaev identity:
\begin{align*}\label{eqn:Pohozaev}
\left(d-6\right)\Vert f' \Vert^2 +(d+2) \Vert r f \Vert^2=\omega(d-2)\Vert f \Vert^2.
\end{align*}
It follows that for $d\geq 6$ it holds $\omega\geq 0$, giving us a range of possible ground state frequencies $\omega\in[0,d]$. Numerical results (Fig.\ \ref{fig:plotd}) show that critical case saturates this range. The key assumption in this reasoning is $d\geq 6$. In fact, in subcritical dimensions this result does not hold and for every $\omega<d$ there exists a ground state \cite{Cao12,Fen16,Mor17}.

Interestingly, for $d>6$ we may strengthen the lower limit on $\omega$ even further. By keeping $ \int_0^\infty f(r)^2 v(r) r^{d-1}  \, dr$ and removing $\Vert f' \Vert^2$ from Eqs.\ (\ref{eqn:Poh1}) -- (\ref{eqn:Poh4}) we obtain the alternative Pohozaev identity:
\begin{align}\label{eqn:Pohozaev2}
8\Vert r f \Vert^2 -(d-6)\int_0^\infty f(r)^2 v(r) r^{d-1} \, dr =4\,\omega\Vert f \Vert^2.
\end{align}
Let us recall that $d$ is the lowest eigenvalue of a linear operator $-\Delta+r^2$ (realised by the eigenfunction $e_0$). Hence, by expressing function $f$ in a base of this operator eigenstates and then using Eq.\ (\ref{eqn:Poh1}) we obtain
\begin{align*}
d \Vert f \Vert^2 \leq \Vert f' \Vert^2 + \Vert r f \Vert^2 = \omega \Vert f \Vert^2 -  \int_0^\infty f(r)^2 v(r) r^{d-1} \, dr.
\end{align*}
Now we may get rid of the last integral with Eq.\ (\ref{eqn:Pohozaev2}):
\begin{align*}
d \Vert f \Vert^2 \leq  \omega \Vert f \Vert^2 +\frac{4\omega}{d-6}  \Vert f \Vert^2- \frac{8}{d-6} \Vert r f \Vert^2.
\end{align*}
Eventually it gives us
\begin{align*}
\omega \geq d- \frac{4d}{d-2}+\frac{8}{d-2} \frac{ \Vert r f \Vert^2}{ \Vert f \Vert^2},
\end{align*}
so for $d>6$ we get an improved lower bounds: $\omega\in\left[\frac{d-6}{d-2}\, d,d\right]$. From the derivation it is clear that they are not optimal.

\subsection{Continuity}
One may show continuity of $\omega(b)$ pretty easily, using some of the already mentioned tools. Let us consider the space $(b,\omega)\in\Lambda :=\mathbb{R}_+ \times [0,d]$, where $\mathbb{R}_+=(0,\infty)$ is the open interval. All values of $b$ and $\omega$ from this set compose a valid pair of initial condition and parameter for Eq.\ (\ref{eqn:SNHlrad}) to have a locally defined solution. Hence, we may decompose set $\Lambda$ into a disjoint union of three sets:
\begin{align*}
J_+=\{&(b,\omega)\in\Lambda \, | \, \exists r_0>0 : f'(r_0)=0\\
&\mbox{ while }f(r)>0 \mbox{ for } r\in(0,r_0)\},\\
J_-=\{&(b,\omega)\in\Lambda \, | \, \exists r_0>0 : f(r_0)=0\\
&\mbox{ while }f'(r)<0 \mbox{ for } r\in(0,r_0)\},\\
J_0=\{&(b,\omega)\in\Lambda \, | \, f(r)>0\mbox{ and } f'(r)<0 \mbox{ for all } r>0\}.
\end{align*}
Continuous dependence on the initial condition and parameter means that sets $J_+$ and $J_-$ are open in $\Lambda$, hence $J_0$ is closed as a complement of $J_+\cup J_-$. We already know that not only for each $b>0$ there exists exactly one value of $\omega$ such that $(b,\omega)\in J_0$, but also such $\omega\in[0,d]$.  Hence, $J_0$ is a graph of $\omega(b)$ function in $\Lambda$. It means that $\omega(b)$ as a function from $\mathbb{R}_+$ to $[0,d]$ is continuous since its graph is closed and codomain is compact.

\subsection{Small $b$ behaviour}\label{sec:bifurcation}
For $b=0$ there exists a unique trivial solution $f\equiv 0$, regardless of the choice of $\omega$. From this line in $(b,\omega)$ plane there bifurcates a branch of our ground states $\omega(b)$. To show this and investigate the initial shape of this branch it is more convenient to consider SNH system in the form of Eq.\ (\ref{eqn:SNHnl}). Then we may define a functional $\mathcal{F}$
\begin{align*}
\mathcal{F}(\omega,f)= -\Delta f+|\mathbf{x}|^2 f-A_d\left(\int_{\mathbb{R}^d} \frac{|f(\mathbf{y})|^2}{|\mathbf{x}-\mathbf{y}|^{d-2}}\, d^d y\right) f-\omega f.
\end{align*}
It satisfies $\mathcal{F}(\omega,0)\equiv 0$ and $\mathcal{F}_f(\omega,0)[u]=-\Delta u+ |\mathbf{x}|^2 u -\omega u$. For $\mathcal{F}_f(\omega,0)$ to be non-invertable, $\omega$ has to be an eigenvalue of this linear operator (i.e.\ $\omega=d+4n$ with $n\in\mathbb{N}$). As ground states have frequency $\omega\in[0,d]$, let us fix $\omega=d$ as the only admissible bifurcation point with $e_0$ being an eigenfunction of $\mathcal{F}_f(d,0)$. At this point we have
\begin{align*}
\mathcal{F}_{\omega,f}(d,0)[u]&=-u,\\
\mathcal{F}_{f,f}(d,0)[u]^2&=0,\\
\mathcal{F}_{f,f,f}(d,0)[u]^3&=-6A_d\left(\int_{\mathbb{R}^d} \frac{|u(\mathbf{y})|^2}{|\mathbf{x}-\mathbf{y}|^{d-2}}\, d^d y\right) u.
\end{align*}
The standard local bifurcation theory \cite{Amb93} thus gives us a subcritical bifurcation with solutions of Eq.\ (\ref{eqn:SNHnl}) given by
\begin{equation}\label{eqn:bifur0}
u=\pm\left(6(d-\omega)\frac{(e_0,\mathcal{F}_{\omega,f}(d,0)[e_0])}{(e_0,\mathcal{F}_{f,f,f}(d,0)[e_0]^3)}\right)^{1/2}+\mathcal{O}(|\omega-d|)
\end{equation}
for $\omega$ values close to $d$. As we are interested in positive solutions, we focus on the branch with plus sign. Obviously $(e_0,\mathcal{F}_{\omega,f}(d,0)[e_0])=-1$, while with the use of the Newton formula (\ref{eqn:newton}) one gets
\begin{align*}
(e_0,\mathcal{F}_{f,f,f}(d,0)[e_0]^3)&=-\frac{6}{2^{\frac{d}{2}-1}(d-2)\Gamma(d/2)},
\end{align*}
concluding in
\begin{equation*}\label{eqn:bifur1}
u=\sqrt{2^{\frac{d}{2}-1}(d-2)\Gamma(d/2)}\left(d-\omega\right)^{1/2}e_0+\mathcal{O}(|\omega-d|).
\end{equation*}
As $u(0)=b$, this result gives us the explicit expression for $\omega(b)$ in the limit of small $b$:
\begin{equation}\label{eqn:bifur}
\omega(b)=d-\frac{b^2}{2^{d/2}(d-2)}+\mathcal{O}(b^3).
\end{equation}
We show this curve in Fig.\ \ref{fig:plotapprox}.

\subsection{Large $b$ behaviour}
\begin{figure}
\centering
\includegraphics[width=0.45\textwidth]{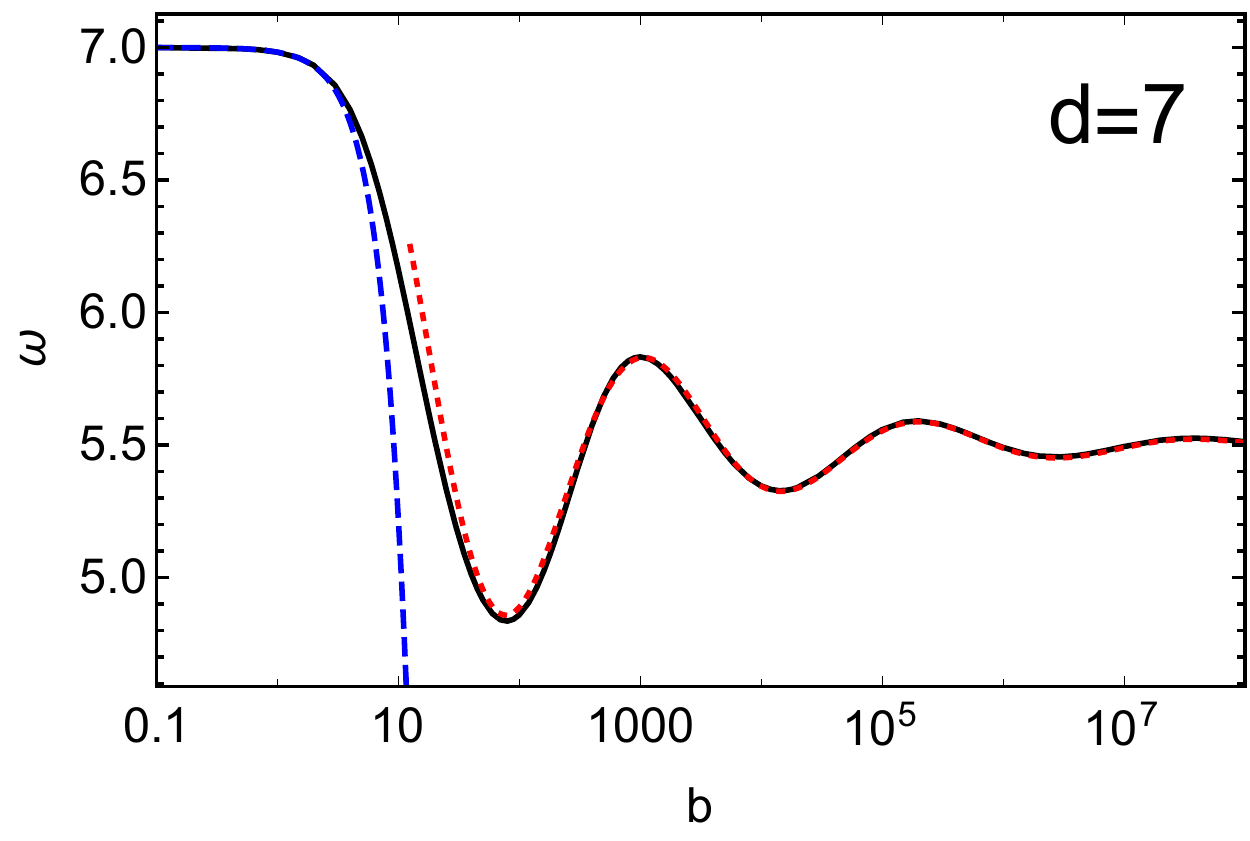}
\caption{Relation between $b$ and $\omega$ for ground state of SNH in supercritical case $d=7$ compared with the approximations: the blue dashed line is given by Eq.\ (\ref{eqn:bifur}) while the red dotted line comes from Eq.\ (\ref{eqn:binf7}).}
\label{fig:plotapprox}
\end{figure}

We consider again Eq.\ (\ref{eqn:binf1}) and reduce it to an autonomous equation with the use of the Emden-Fowler transformation, $s=\ln \rho$, $g(s)=\rho^2 F(\rho)$, obtaining:
\begin{align*}\label{eqn:binf2}
g'' + (d-6) g' -2(d-4)g+ g^2&= 0.
\end{align*}
This equation can be investigated with dynamical systems methods. It has a nontrivial fixed point $g=2(d-4)$, corresponding to the solution of Eq.\ (\ref{eqn:binf1}): $F(\rho)=2(d-4)/\rho^2$. Linearisation around it (i.e., employing $g=2(d-4)+\nu$ and preserving only terms linear in $\nu$) gives
\begin{align}\label{eqn:binf3}
\nu'' + (d-6) \nu' +2(d-4)\nu&= 0.
\end{align}
Reintroducing quantities $\beta=-\frac{d}{2}+3$ and $ \alpha_1=\frac{1}{2}\sqrt{|d^2-20d+68|}$, this linear system (\ref{eqn:binf3}) has for $7\leq d\leq 15$ eigenvalues of a form $\beta\pm i \alpha_1$, while for $d\geq 16$ of $\beta\pm \alpha_1$. This change of nature of eigenvalues at $d=10+4\sqrt{2}\approx 15.66$ carries the change of behaviour of solutions with large $b$ when dimension $d$ changes from 15 to 16. Indeed, for $7\leq d\leq 15$ keeping the leading terms yields
\begin{align*}\label{eqn:binf4}
g(s)\approx2(d-4)\left[1+A e^{\beta s}\sin(\alpha_1 s+\delta) \right],
\end{align*}
where $A$ and $\delta$ are some constants. Going back to the original variables we get
\begin{align*}\label{eqn:binf5}
f(r)\approx\frac{2(d-4)}{r^2}\left[1+A (\sqrt{b}r)^\beta \sin\left(\alpha_1\ln \sqrt{b} r+\delta\right) \right].
\end{align*}
This approximation is valid in an intermediate range $1/b \ll r \ll b$. On the other hand, for large values of $r$, when the harmonic term dominates, $f$ behaves like a solution of a linear harmonic oscillator
\begin{equation*}\label{eqn:binf6}
	f(r)\approx C\, e^{-r^2/2}\, U\left(\frac{d-\omega}{4},\frac{d}{2},r^2\right),
\end{equation*}
with $C$ being some constant and $U$ denoting the confluent hypergeometric function of the second kind. We may consider some adequately large values of $b$ and fix some $r_0$ such that in this point both approximations apply. Then we get two expressions for $f(r_0)$, one depending on $b$ directly, the second one through $\omega$. For large $b$ value we may expand the second one into $f(r_0)\approx C_0+C_1(\omega-\omega_\infty)$, where $C_0$ and $C_1$ are some constants. Comparing non-constant terms in both expressions we get
\begin{align}\label{eqn:binf7}
\omega(b)\approx \omega_\infty + \tilde{A} b^{\beta/2} \sin\left(\alpha_1\ln \sqrt{b} +\tilde{\delta} \right).
\end{align}
We compare this approximation with exact numerical results in Fig.\ (\ref{fig:plotapprox}).

From these considerations it is evident that for $d\geq 16$ these oscillations vanish and function $\omega(b)$ becomes monotone (c.f.\ Fig.\ \ref{fig:plotd}). Analogous observation was made for a Gross-Pitaevskii equation, when the dimension changes from 12 to 13 \cite{Bizip}.

\section{Summary}\label{sec:summary}
In this article, motivated by connections to AdS stability problem, we started investigations of the Schr\"{o}dinger-Newton-Hooke equation in supercritical dimensions ($d > 6$). As a first step we concentrated on spherically symmetric stationary solutions with the main focus on the ground state. Instead of usually considered nonlocal Eq.\ (\ref{eqn:SNHnlrad}) \cite{Cao12, Car05, Che17, Che18, Fen16, Fro03, Luo19, Sch17, Wan08}, our description was based on the equivalent system (\ref{eqn:SNHlrad}), which let us to use typical tools from the theory of ordinary differential equations and dynamical systems, especially the shooting method. With this method we proved existence of a whole ladder of solutions characterised by the number of zeroes for any positive central field value $b$. We also showed that for a fixed $b$, the ground state is unique, which allowed us to define its frequency as $\omega(b)$. We investigated some properties of this function in various dimensions, in particular showed its continuity and restricted its possible values. We also studied its behaviour for small $b$, when the solutions bifurcate from the linear quantum oscillator ground states, and for large $b$, when the solutions tend to the singular solutions. It turned out, that the behaviour of $\omega(b)$ is different for $7\leq d\leq 15$, when it is an oscillating function, than for $d\geq16$, when it becomes monotonically decreasing. 

A quick look into literature reveals that some features of SNH described here (including the change of shape of $\omega(b)$ in higher dimensions) are shared by various different quasilinear problems with confinement in their respective supercritical dimensions. Examples are mostly restricted to systems confined in the ball-shaped domains with no potential and include Gross-Pitaevskii equation \cite{Bud87, Dol07, Guo11, Miy14} and Gelfand problem \cite{Jos73}. To the best of our knowledge, except this work, the only results regarding unbounded systems with confinement achieved by the presence of an external potential appeared in Refs.\  \cite{Bizip, Sel11, Sel12, Sel13} and concerned Gross-Pitevskii equation with harmonic trap. Many similarities between these results suggest that there is a common framework able to describe these behaviours.

Going back to the main motivation of this paper, results covered here are just a starting point in the further work into the understanding of supercritical SNH system in connection with the AdS stability problem. In the sequel to this work we plan to pursue this path by investigating stability of stationary solutions found here and by looking into dynamics of this system.

\section*{Acknowledgements}
I am very thankful to Piotr Bizo\'n for his guidence, help and many revisions of this work. I would also like to thank Dmitry E. Pelinovsky for his remarks to the manuscript. Together with Szymon Sobieszek, he shared with me many inspiring ideas during my short stay at McMaster University. Finally, I acknowledge hospitality and support showed by the Mittag-Leffler Institute within the General Relativity, Geometry and Analysis: beyond the first 100 years after Einstein program. This project was funded by the Polish National Science Centre Grant No. 2020/36/T/ST2/00323 and Grant No. 2017/26/A/ST2/00530.

\end{document}